\documentstyle[12pt]{article}
\begin{document}
\begin{center}
{\bf HADRON MULTIPLICITIES AT THE ENERGIES}\\
{\bf OF LEP-1.5 AND LEP-2}\\
\vspace*{.3cm}
A.V. Kisselev, V.A. Petrov\\
{\small Institute for High Energy Physics}\\
{\small 142284 Protvino (Moscow Region), Russia}
\end{center}
\begin{quote}
Total hadron multiplicities and multiplicities of hadrons in events
with heavy quarks in $e^+e^-$ annihilation at the energies of LEP-1.5 
and LEP-2 are calculated on the basis of QCD.
\end{quote}

One of the most important overall characteristics of final hadronic states in
$e^+e^-$ annihilation is an average multiplicity of (charged) hadrons,
$< n >_{had}$. The rise of $< n >_{had}$ in $W$ 
enables one to make conclusions on a mechanism of multiple hadron production 
in hard processes.

The data on $< n >_{had}(W)$ with the high energy data from
LEP-1~\cite{LEP1} and SLC~\cite{SLC} included are well approximated by the
QCD--based expressions (see, for instance,~\cite{Giacomelli}). Let us remind 
that at $W=m_Z$ the world average is equal to
\begin{equation}
< n >_{had} = 20.94 \pm 0.20. \label{1}
\end{equation}

The total multiplicity in $e^+e^-$ annihilation is given by the formula:
\begin{equation}
< n >_{had} \sum_q P_q = \sum_q < n >_q P_q,
\label{2}
\end{equation}
where $P_q$ is a SM weight of an event with primary quarks of type $q$ 
($q=u,d,s,c,b$) and $< n >_q$ is an average multiplicity in
such an event.

In Ref. \cite{Petrov} in the framework of QCD we have derived the following
expression for $< n >_q$ (see~\cite{Petrov} for details):
\begin{equation}
< n >_q = < n >_q^0 + C_F \int \frac{dk^2}{k^2}
\frac{\alpha_s(k^2)}{\pi} E(W^2,k^2) n_g(k^2). \label{3} 
\end{equation}
Here $< n >_q^0$ means the average multiplicity of fragmentation
products
of the primary quark $q$. $E(W^2,k^2)$ describes an inclusive distribution of
gluon jets in their invariant masses $k^2$, while $n_g(k^2)$ is a hadron 
multiplicity inside the gluon jet with the virtuality $k^2$.

Formulae (\ref{2}), (\ref{3}) made it possible to describe well hadron
multiplicities  in $e^+e^-$--events induced by $b$--quarks, 
$< n >_b$,~\cite{DELPHI,OPAL} and to predict the values of hadron 
multiplicities in events induced by $c$--ª¢ àª ¬¨, $< n >_c$.
In particular, our result, $(< n >_c - 
< n >_{uds})(m_Z) = 1.01$, where $< n >_{uds}$ means
an average multiplicity of hadrons in events with light primary quarks,
have appeared to be in good agreement with the data from OPAL and SLD
Collaborations~\cite{OPAL,SLD}.

In the present paper we calculate hadron 
multiplicities at the energies of LEP-1.5 (133 GeV) and LEP-2 (161, 175, 
192 GeV), with the use of Eqs.~(\ref{2}), (\ref{3}).
The results for $m_c=1.5$ GeV/c$^2$, $m_b=4.8$ GeV/c$^2$ are presented in the
Table.

\begin{center}
\begin{tabular}{||l|c|c|c|c||}
\hline
$W$, Ē' & 133 & 161 & 175 & 192 \\ \hline
$< n >_{uds}$ & 23.13 & 25.02 & 25.87 & 26.85 \\ \hline
$< n >_c$ & 24.13 & 26.02 & 26.88 & 27.85 \\ \hline
$< n >_b$ & 26.80 & 28.65 & 29.54 & 30.51 \\ \hline
$< n >_{had}$ & 24.10 & 26.00 & 26.85 & 27.82 \\ 
\hline
\end{tabular}
\end{center}

Recently the data on total multiplicity at $W=130$~GeV and 
$W=133$~GeV have been obtained~\cite{LEP1.5D,LEP1.5O}:
\begin{eqnarray}
\mbox{\rm DELPHI} (W=130 \mbox{\rm GeV}): \ < n >_{had} 
&=& 23.84 \pm 0.51 \pm 0.52, \nonumber \\
\mbox{\rm OPAL} (W=133 \mbox{\rm GeV}):\hphantom{HI}\ < n 
>_{had} &=& 23.40 \pm 0.45 \pm 0.47.\label{4}
\end{eqnarray}
In Ref. \cite{Giacomelli} the corrected data at $W=133$ GeV are presented:
\begin{eqnarray}
\mbox{\rm DELPHI:} \ < n >_{had} &=& 23.3 \pm 0.6, \nonumber \\
\mbox{\rm OPAL:\hphantom{HI}} \ < n >_{had} &=& 23.24 \pm 0.32 
\pm 0.41.\label{5}
\end{eqnarray}

Finally, there are LEP-2 data on the total multiplicity~\cite{LEP2}:
\begin{equation}
< n >_{had}(161 \mbox{\rm GeV}) = 25.78 \pm 0.45 \pm 0.53.
\label{6}
\end{equation}

As one can see from the Table, our values $< n >_{had}(133 
\mbox{\rm GeV})=24.10$ and $< n >_{had}(161 \mbox{\rm GeV})=
26.00$ agree well with the data~(\ref{4})-(\ref{6}). 
Note, that Monte Carlo models accounting for the quark masses give the 
values $24.1 \div 24.2$ at $W=133$~GeV~\cite{LEP1.5O} which are very closed 
to our prediction.

At the same time, empirical fits and QCD--formulae which do not take into 
account the specific features of the events with the heavy quarks 
(see, for instance,~\cite{Webber}), result in somewhat higher value
$< n >_{had}=24.4$ at $W=133$ GeV~\cite{Giacomelli,LEP1.5O}.

These discrepancies can be, of course, "improved" provided one adds in a fit 
the value of $< n >_{had}$ at the point $W=133$ GeV. 
It is clear, however, that such an "improvement" could be hardly considered
to be satisfactory as it actually lowers predictive power of a theory.

Our results for $W=175$ GeV (see Table) can be compared, for instance, with
those in Ref.~\cite{Knowles}, where the values 
$< n >_c=28.8$, $< n >_b=30.6$, 
$< n >_{had}=27.0$ are obtained and with the value
$< n >_{had}=27.3$ in Ref.~\cite{Ybook}.

\vfill \eject

\begin{thebibliography}{99}
\bibitem{LEP1}
P. Abreu, W. Adam, F. Adami et al. (DELPHI Coll.), Z. Phys. {\bf C50}, 
185 (1991).
B. Aveda, O. Adriani, M. Aguilar-Benitez et al. (L3 Coll.), Phys. Lett. 
{\bf B259}, 199 (1991).
D. Decamp, B. Deschizeaux, C. Goy et al. (ALEPH Coll.), Phys. Lett. 
{\bf B273}, 181 (1991).
P.D. Acton, G. Alexander, J. Allison et al. (OPAL Coll.), Z. Phys. 
{\bf C53}, 539 (1992).
\bibitem{SLC}
G.S. Abrams, C.E. Adolphsen, D. Averill et al. (MARKII Coll.), Phys. Rev. 
Lett. {\bf 64}, 1334 (1990).
K. Abe, I. Abt, W.W. Ash et al. (SLD Coll.), Phys. Rev. Lett. 
{\bf 72}, 3145 (1994).
\bibitem{Giacomelli}
G. Giacomelli and P. Giacomelli, Preprint DFUB 96/10, Bologna, 1996.
\bibitem{Petrov}
V.A. Petrov and V.A. Kisselev, Z. Phys. {\bf C66}, 453 (1995).
Nucl. Phys. B (Proc. Suppl.) {\bf 39B, C}, 364 (1995).
\bibitem{DELPHI}
P. Abreu, W. Adam, T. Adye et al. (DELPHI Coll.), Phys. Lett. 
{\bf B347}, 447 (1995).
\bibitem{OPAL}
R. Akers, G. Alexander, J. Allison et al. (OPAL Coll.), Phys. Lett. 
{\bf B352}, 176 (1995).
\bibitem{SLD}
K. Abe, K. Abe, I. Abt et al. (SLD Coll.), Phys. Lett. {\bf B386}, 475 (1996).
\bibitem{LEP1.5D}
P. Abreu, W. Adam, T. Adye et al. (DELPHI Coll.), Phys. Lett. 
{\bf B372}, 172 (1996).
\bibitem{LEP1.5O}
G. Alexander, J. Allison, N. Altekamp et al. (OPAL Coll.), 
Preprint CERN-PPE/96-47, 1996.
\bibitem{LEP2}
P.Abreu, A. De Angelis, R. Henriques and M. Pimenta, DELPHI Note 96-144
Phys. 642, 1996.
\bibitem{Webber}
B.R. Webber, Phys. Lett. {\bf B143}, 501 (1984).
\bibitem{Knowles}
I. Knowles, {\em in Proc. of the EPS Conference on High Energy Physics}, 
Eds. J. Lemonne, C. Vander Velde and F. Verbeure, Brusseles, 1995, p.~349.
\bibitem{Ybook}
{\em Physics at LEP2}, Eds. G. Altarelli, T. Sj\"ostrand and 
F. Zwirner, Geneva, 1996, v.~1, p.~276.
\end{thebibliography}
\end{document}